\documentclass{elsart}

\usepackage{natbib}

 \usepackage{epsfig}

\usepackage{amssymb}

\def\arcsec{\hbox{$^{\prime\prime}$}}

\begin{document}

\begin{frontmatter}

% Title, authors and addresses

% use the thanksref command within \title, \author or \address for footnotes;
% use the corauthref command within \author for corresponding author footnotes;
% use the ead command for the email address,
% and the form \ead[url] for the home page:
% \title{Title\thanksref{label1}}
% \thanks[label1]{}
% \author{Name\corauthref{cor1}\thanksref{label2}}
% \ead{email address}
% \ead[url]{home page}
% \thanks[label2]{}
% \corauth[cor1]{}
% \address{Address\thanksref{label3}}
% \thanks[label3]{}

\title{New Detections of Optical Emission from Kiloparsec-scale Quasar 
Jets\thanksref{label1}}
\thanks[label1]{Radio astronomy at Brandeis University is supported by 
the National Science Foundation through grants AST 98-02708 and AST 
00-98608.}

% use optional labels to link authors explicitly to addresses:
% \author[label1,label2]{}
% \address[label1]{}
% \address[label2]{}

\author{C.~C. Cheung, J.~F.~C. Wardle, Tingdong Chen \& S.~P. Hariton}

\address{Department of Physics, MS~057, Brandeis University, Waltham, MA
02454, USA}

\begin{abstract}

We report initial results from the detection of optical emission in the
arcsecond-scale radio jets of two quasars utilizing images from the {\it
Hubble Space Telescope} archive. The optical emission has a very knotty
appearance and is consistent with synchrotron emission from highly
relativistic electrons in the jet. Combining these observations with those
of previously reported features in other quasars, an emerging trend
appears to be that their radio-to-optical spectral indices are steeper
than those of similar features in jets of lower power radio sources.

\end{abstract}


\end{frontmatter}

% main text
\section{Motivation}
\label{}

The number of active galaxies observed to have radio-bright kpc-scale
relativistic jets emanating from their nuclei has grown to many 100's over
the last several decades \citep{bri84,liu02}, but only a handful of them
were known to have X-ray counterparts prior to the launch of the {\it
Chandra} X-ray Observatory.  {\it Chandra} has in just the last three
years netted X-ray detections of jet features in over 30 more radio
sources \citep[see][and the most recent updates in the associated website
-- http://hea-www.harvard.edu/XJET/]{harr02}, and many more are being
discovered as a result of dedicated searches (see e.g.  individual
contributions by R.~M. Sambruna, H.~L. Marshall, and D.~A. Schwartz).  
With the increased interest in the X-rays, studies in the optical/UV
regime have also enjoyed a resurgence of activity. This is because the
optical is the only other accessible window in the electromagnetic
spectrum where comparable sub-arcsecond resolution can be readily achieved
with the {\it Hubble Space Telescope}\footnote{Based on observations made
with the NASA/ESA Hubble Space Telescope, obtained from the data archive
at the Space Telescope Science Institute. STScI is operated by the
Association of Universities for Research in Astronomy, Inc. under NASA
contract NAS 5-26555.}, and bridges the $\sim$6-7 orders of magnitude gap
in frequency coverage between radio and X-ray studies of resolved
features. It turns out that the optical fluxes and spectra can be a key
diagnostic in distinguishing between different X-ray production
mechanisms.  Specifically, the faintness of the optical knots in the
`first-light' X-ray jet detection in the powerful quasar PKS~0637--752
\citep{cha00,sch00} helped to rule out canonical synchrotron and
synchrotron self-Compton models, and established the importance of X-rays
produced via inverse Compton scattering of CMB photons by synchrotron
emitting electrons in the kpc-scale jets \citep{tav00,cel01}. Also, if the
emission mechanism responsible for the optical emission is synchrotron
radiation, as is commonly accepted, observations in this waveband probe
the physics of the most energetic and short-lived electrons, and are key
to studying acceleration processes in the jet.

In the pre-{\it Chandra} era, optical jets were predominantly found in low
power sources with the {\it HST} \citep[e.g.,][]{spa94,sca02}, and optical
counterparts to jets in quasars were difficult to detect with ground based
facilities. Only a handful of examples existed, courtesy of the {\it HST}
\citep[e.g.,][]{rid97}, so optical jets in quasars were thought to be
rare. However, {\it Chandra's} successful X-ray detections of quasar jets
has forced observers to examine new and archival {\it HST} data more
carefully, resulting in many new detections in the optical.  This resulted
in the identifications of a number of bright optical knots coincident with
peaks in the radio jets of about ten quasars in a recent {\it Chandra/HST}
cycle-2 search for such emission \citep[][R. Scarpa et al., in
prep.]{sam02}. We applied the same techniques employed in this survey and
found a bright optical/UV knot in the kpc-scale radio jet of 3C~279 in
archival {\it HST} data \citep{che02} which encouraged us to search for
more such emission in other available archival data.

\section{New Detections and Search Strategy}

Figures~1~\&~2 show two of our first three optical detections in the
quasars 3C~454.3 and PKS~1229--021 as a result of our search.  The third
detection is in the bright radio hotspot in the lobe-dominated quasar
3C~263 which was recently published by \citet{har02} along with the {\it
Chandra} X-ray detection of the feature. These quasars were known to us to
exhibit bright radio jets and were obvious candidates to search for
optical counterparts. The optical emission in these and other known
kpc-scale quasar jets \citep[][R. Scarpa et al., in
prep.]{rid97,cha00,sam02,che02} is compact, match peaks in the radio jets
well, and is consistent with synchrotron emission from highly relativistic
electrons in the jet.

\begin{figure*}
\hrule
%\vspace*{8cm}
\vbox to 7cm{\vfill\hbox to \hsize{\hfill 
\includegraphics*[width=0.5\textwidth,angle=270]{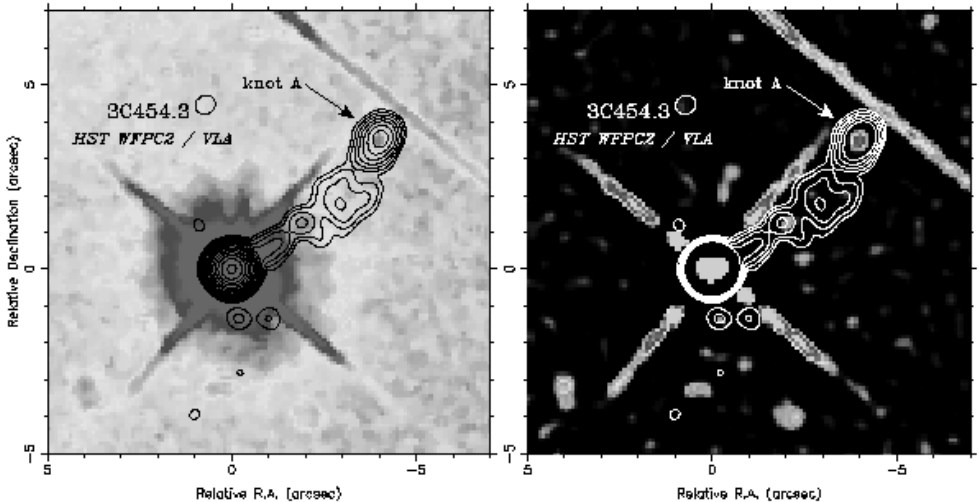}\hfill}\vfill}
\hrule
\caption{{\it HST} WFPC2 F702W grayscale images of 3C~454.3
($z=0.859$) with our new VLA 8.4 GHz 0.5\arcsec~resolution image of the 
radio jet overlaid.  Radio contours increase by factors of two from 
1 mJy/beam (only the first 5 levels are plotted in the right panel).  
The {\it HST} data are shown before (left) and after
(right) applying an unsharp masked filter in order to suppress emission
from the optical nucleus.  A sharp optical feature clearly lies coincident
with the terminal radio knot A about 5.3\arcsec~(44 kpc projected) 
distant.  The four diffraction spikes around the optical nucleus
are clearly present, as is a portion of a spike in the upper right of 
the displayed field of view from a field star.}
\end{figure*}

\begin{figure*}
\hrule
%\vspace*{8cm}
\vbox to 6.5cm{\vfill\hbox to \hsize{\hfill 
\includegraphics*[width=0.45\textwidth,angle=270]{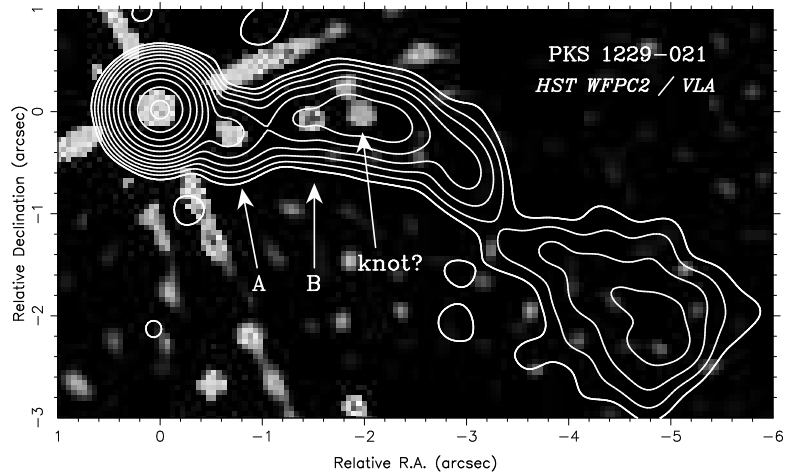}\hfill}\vfill}
\hrule
\caption{{\it HST} WFPC2 F702W grayscale image of PKS~1229--021
($z=1.045$) with an unsharp masked filter applied (as in the right panel
of Fig.~1) and a VLA 15 GHz image of the jet (reprocessed from data
published in Kronberg et al., 1992) at 0.4\arcsec~resolution overlaid.  
Radio contours are spaced by factors of 2 from 0.3 mJy/beam.
We found at least two bright optical knots following the first
two peaks in the large-scale radio jet (confirmed also in an 
archival F450W image).}
\end{figure*}

With the growing number of optical jet detections in radio-loud sources,
we can begin to make comparisons between the jet properties of different
subclasses. One emerging trend is that the radio-to-optical spectral
indices of the jet features in quasars (see above references) appear to be
steeper by $\Delta\alpha\sim0.2$ than those measured in lower power radio
sources \citep[see e.g.,][]{spa94,sca02}.

After these initial detections, we decided on a more systematic approach
to mining through the available archival data. We are currently searching
the {\it HST} archive database for images of radio bright ($>$ 0.5 Jy at 5
GHz) active galaxies obtained after $\sim$1994 (post-COSTAR), mostly
concentrating on WFPC2 images which have $\geq$ 1 orbit of total exposure
time. The resultant list of candidates are being correlated with the
compilation of \citet{liu02} of sources known to exhibit radio jets. As a
result of this ongoing effort, we have found optical counterparts to knots
in an additional three quasar jets and have already obtained archival
multi-frequency MERLIN and VLA data for our analysis.  These results will
be presented along with the complete analysis of the detections presented
here in a forthcoming paper.


\begin{thebibliography}{}

% \bibitem[Names(Year)]{label} or \bibitem[Names(Year)Long names]{label}.
% (\harvarditem{Name}{Year}{label} is also supported.)

\bibitem[Bridle \& Perley(1984)]{bri84}
Bridle, A.~H. \& Perley, R.~A. 1984, ARA\&A, 22, 319 

\bibitem[Celotti, Ghisellini, \& Chiaberge(2001)]{cel01} Celotti, A.,
Ghisellini, G., \& Chiaberge, M.\ 2001, MNRAS, 321, L1

\bibitem[Chartas et al.(2000)]{cha00} Chartas, G.~et al.\ 2000, ApJ, 542,
655

\bibitem[Cheung(2002)]{che02} Cheung, C.~C.\ 2002, ApJL, 581, L15

%\bibitem[Fraix-Burnet et al.(1991)]{fra91} Fraix-Burnet, D. et al.\ 1991, 
%AJ, 101, 88

%\bibitem[Fraix-Burnet et al.(1991)]{fra91} Fraix-Burnet, D., Golombek,
%D., Macchetto, F., Nieto, J.-L., Lelievre, G., Perryman, M.~A.~C., \& di
%Serego Alighieri, S.\ 1991, AJ, 101, 88

\bibitem[Hardcastle et al.(2002)]{har02} Hardcastle, M.~J. et al.\ 2002,
ApJ, accepted (astro-ph/0208204)

%\bibitem[Hardcastle et al.(2002)]{har02} Hardcastle, M.~J., Birkinshaw,  
%M., Cameron, R.~A., Harris, D.~E., Looney, L.~W.,\& Worrall, D.~M.\ 2002,
%ApJ, accepted (astro-ph/0208204)

\bibitem[Harris \& Krawczynski(2002)]{harr02} Harris, D.~E.~\& 
Krawczynski, H.\ 2002, ApJ, 565, 244

\bibitem[Kronberg, Perry, \& Zukowski(1992)]{kro92} Kronberg, P.~P.,
Perry, J.~J., \& Zukowski, E.~L.~H.\ 1992, ApJ, 387, 528

\bibitem[Liu \& Zhang(2002)]{liu02} Liu, F.~K.~\& Zhang, Y.~H.\ 2002,
A\&A, 381, 757

%\bibitem[Mukherjee et al.(1997)]{muk97} Mukherjee, R., Bertsch, D.~L.,
%Bloom, S.~D.~et al.\ 1997, ApJ, 490, 116

%\bibitem[O'Dea et al.(1999)]{ode99} 
%O'Dea, C.~P., de Vries, W., Biretta, J.~A., \& Baum, S.~A.\ 1999, AJ, 
%117,  1143 

%\bibitem[Parma et al.(2002)]{par02} Parma, P. et al.\ 2002, A\&A, 
%accepted (astro-ph/0210461)

%\bibitem[Parma et al.(2002)]{par02} Parma, P., de Ruiter,
%H.~R., Capetti, A., Fanti, R., Morganti, R., Bondi, M., Laing, R.~A., \& 
%Canvin, J.~R.\ 2002, A\&A, accepted (astro-ph/0210461)

\bibitem[Ridgway \& Stockton(1997)]{rid97} Ridgway, S.~E.~\& Stockton, A.\
1997, AJ, 114, 511

\bibitem[Sambruna et al.(2002)]{sam02} Sambruna, R.~M. et al.\ 2002, ApJ, 
571, 206

%\bibitem[Sambruna et al.(2002)]{sam02} Sambruna, R.~M., Maraschi, L., 
%Tavecchio, F., Urry, C.~M., Cheung, C.~C., Chartas, G., Scarpa, R., \&
%Gambill, J.~K.\ 2002, ApJ, 571, 206

\bibitem[Scarpa \& Urry(2002)]{sca02} Scarpa, R.~\& Urry, C.~M.\ 2002, New
Astronomy Review, 46, 405

\bibitem[Schwartz et al.(2000)]{sch00} Schwartz, D.~A., Marshall, H.~L.,
Lovell, J.~E.~J.~et al.\ 2000, ApJL, 540, L69

\bibitem[Sparks, Biretta, \& Macchetto(1994)]{spa94} Sparks, W.~B.,
Biretta, J.~A., \& Macchetto, F.\ 1994, ApJS, 90, 909

\bibitem[Tavecchio et al.(2000)]{tav00} Tavecchio, F. et al.\ 2000, ApJL,
544, L23

%\bibitem[Tavecchio et al.(2000)]{tav00} Tavecchio, F., Maraschi, L.,
%Sambruna, R.~M., \& Urry, C.~M.\ 2000, ApJL, 544, L23

\end{thebibliography}
\end{document}